\documentclass[aps,prl,reprint,amsmath,amssymb,floatfix]{revtex4-2}
\usepackage[bookmarks=false,linkcolor=NavyBlue,urlcolor=NavyBlue,colorlinks,citecolor=NavyBlue]{hyperref}
\usepackage{graphicx}
\usepackage{amsmath}
\usepackage[svgnames]{xcolor}
\usepackage{enumerate}
\usepackage[normalem]{ulem}
\let\oldsection\section
\renewcommand{\section}[1]{\emph{#1.---}}

\newcommand{\figref}[2]{\hyperref[#1]{\ref{#1}(#2)}}
\newcommand{\figsref}[2]{\hyperref[#1]{\ref{#1}#2}}
\newcommand{\Weizmann}{Department of Condensed Matter Physics, Weizmann Institute of Science, Rehovot, Israel 7610001}

\begin{document}
\graphicspath{ {./} }
\title{Electron Interference as a Probe of Majorana Zero Modes}
\author{Nadav Drechsler}
\affiliation{\Weizmann}

\author{Omri Lesser}
\affiliation{\Weizmann}

\author{Yuval Oreg}
\affiliation{\Weizmann}

\begin{abstract}

Detecting Majorana zero modes (MZMs) in topological superconductors remains challenging, as localized non-topological states can mimic MZM signatures. Here, we propose electron interferometry by non-local transport measurements as a definitive probe to distinguish MZMs from non-topological states. We develop an analytical minimal model showing that interference via two MZMs exhibits a robust pattern, in contrast to a non-topological system. We then numerically confirm this using various topological superconductor models. We find that MZMs are characterized by an interference pattern that is insensitive to various perturbations, such as electrostatic gate potential, and resilient to disorder. Our proposed interferometry approach offers an experimentally accessible means to detect MZMs, probing their underlying nature through a universal response.
\end{abstract}
\maketitle

\section{Introduction}Topological materials adhere to a bulk-boundary correspondence principle, establishing a direct link between the topological properties of the bulk and the presence of boundary modes. This is manifested by the closing and reopening of the gap of the bulk modes when the system undergoes a topological phase transition. In quasi-one-dimensional topological superconductors (1D-TSCs), this phenomenon is intimately tied to the emergence of Majorana zero-energy modes (MZMs) at the two ends of the TSC. These MZMs are exotic quasi-particles predicted to have non-Abelian exchange statistics. The interest in realizing MZMs has been propelled by the foundational endeavor of uncovering new phases of matter, alongside their potential applications in topological quantum computation~\cite{hasan_colloquium_2010,alicea_new_2012,leijnse_introduction_2012,nayak_non-abelian_2008,oreg_majorana_2020,oreg_helical_2010,lutchyn_majorana_2010}.

The definite detection of MZMs remains a contentious issue in experimental studies. A primary experimental approach has centered on the observation of local zero-bias conductance peaks (ZBCPs)~\cite{mourik_signatures_2012,das_high-efficiency_2012}. However, non-topological MZM-like signatures originating from disorder might closely mimic the local conductance patterns of genuine MZMs, thereby undermining the reliability of ZBCP-based detection of MZMs~\cite{vuik_reproducing_2019,flensberg_engineered_2021,wang_singlet_2022,dvir_realization_2023,yazdani_hunting_2023,tsintzis_roadmap_2023}. Later works employed non-local conductance measurements, enabling the simultaneous observation of the emergence of MZMs and the closing and reopening of the bulk gap signaling the topological phase transition~\cite{microsoft_quantum_inas-hybrid_2023,banerjee_signatures_2023,banerjee_local_2023}. Others probed the non-local relation of a pair of MZMs with interferometry~\cite{fu_electron_2010, drukier_evolution_2018,hell_distinguishing_2018,sugeta_enhanced_2023,aghaee_interferometric_2024}.

Despite the valuable information provided by these interferometry setups, they pose certain constraints. Some are based on tunneling between a pair of MZMs, and hence depend on the coupling between them~\cite{sugeta_enhanced_2023}. As the two MZMs become more localized and far away from each other, this coupling gets exponentially smaller, and therefore the signal is strongly attenuated, making it difficult to perform such measurements in long systems. Other works assumed high charging energy in the superconductor (SC), which is not compatible with a grounded SC~\cite{fu_electron_2010, drukier_evolution_2018,hell_distinguishing_2018,whiticar_coherent_2020}. 

Our approach is predicated on a simple fact: a fundamental difference between MZMs and non-topological MZM-like states lies in the degrees of freedom present at low energies. A pair of MZMs, each residing at an edge of the topological SC, comprises a single fermionic state. On the other hand, when MZM-like localized states appear in non-topological SCs, they originate in two individual localized fermionic states formed by four Majoranas. 

Harnessing this property and inspired by previous Majorana interferometry works, we offer in this manuscript a novel method to conclusively distinguish between MZMs in TSCs and MZM-like states in non-topological grounded SCs. This method utilizes electron interferometry through non-local transport measurement, by tunneling through the two localized states in parallel (thereby \emph{not} measuring their parity). We first introduce an analytical model demonstrating the distinct behavior of a system coupled to two and four Majoranas. While the former topological case displays a robust universal interference pattern under different perturbations, the latter, non-topological case does not. We then propose two similar experiments that reveal this distinction through the variation of the relative input phase and an electrostatic gate, and test them numerically in various TSC realizations with a full tight-binding model.

\section{Minimal model}
\begin{figure*}
\includegraphics[width=\textwidth]{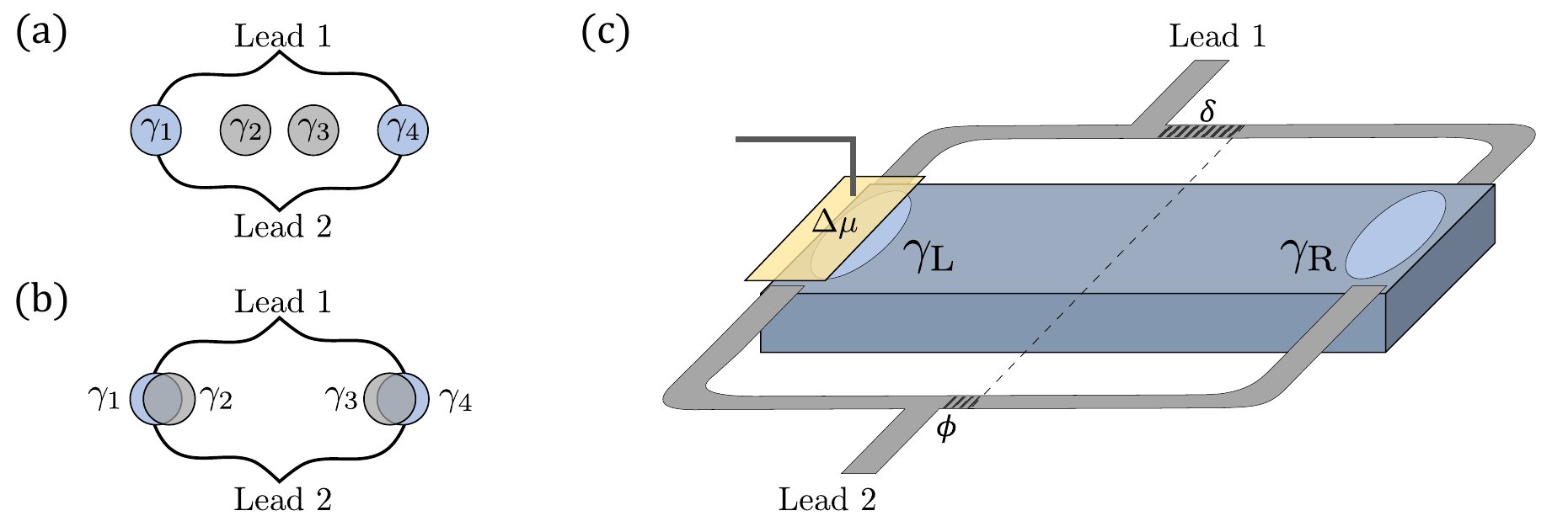}
\caption{Illustration of the experimental scheme expressed in the analytical model. Normal leads are connected to two sides of the system, and electrons tunnel through (a)~two $(\beta=0)$ or (b)~four Majoranas $(\beta\sim 1)$. (c)~Schematic of the experimental setup. Localized states (light blue), from either topological or non-topological origin, reside on both sides of the SC (blue). Four of its corners are connected to lead 1 at the top and lead 2 at the bottom (gray). We calculate the non-local conductance $G$ between lead 1 and lead 2. There are phase differences $\delta$ and $\phi$ between the two arms of leads 1 and 2, correspondingly, either due to magnetic fluxes enclosed between the two arms of each lead or a difference in the lengths of the arms. In the second scheme, an electrostatic gate (yellow) is placed at one side of the SC, locally changing its chemical potential by $\Delta\mu$, covering about $5\%$ of the system's length.\label{fig:system}}
\end{figure*}

We consider a setup of a SC with 4 Majoranas $\left\{ \gamma_{1},\gamma_{2},\gamma_{3},\gamma_{4}\right\}$
that have no matrix elements between one another, so their Hamiltonian $H$ is 
the $4\times4$ zero matrix. The physical system might have additional higher-energy states,
which we ignore in this low-energy treatment. The SC is coupled
to two normal leads: lead 1 (2) is connected to the system's top (bottom) left and
right sides, as shown in Fig.~\ref{fig:system}. For simplicity, we assume there is no reflection in the splitting of each lead to its two arms. This assumption only quantitatively affects the final result, as shown in the numerical results in the Supplemental Material. We calculate the non-local conductance from lead 1 to lead 2.

We first consider the case where each lead is coupled with equal tunneling strength $\Gamma$ to all four Majoranas ${\gamma_1, \gamma_2, \gamma_3, \gamma_4}$, forming a single fermion on each side of the SC, $f_{\rm L(\rm R)}^{\dagger}=\gamma_{1(3)} + i\gamma_{2(4)}$. We introduce a phase difference $\delta$ between the left and right arms of lead 1 originating, for instance, from a magnetic flux enclosed between the two arms of the lead, or from different lengths of the lead's arms. This is described by the tunneling term $\Gamma c_1^{\dagger}\left[\gamma_1 - i\gamma_2 + e^{i\delta}(\gamma_3 - i\gamma_4)\right] + \mathrm{H.c.}$ between lead 1 and the Majoranas, where $c_1^{\left(\dagger\right)}$ is the particle (hole) annihilation
operator in lead 1. We similarly introduce another phase difference $\phi$ between the arms of lead 2 at the output.

To model a difference in the coupling strength of the lead to the inner Majoranas $\gamma_2,\gamma_3$, we define a parameter $\beta$ that quantifies their tunneling amplitude relative to the outer Majoranas $\gamma_1,\gamma_4$. $\beta$ range from 0 to 1, representing no to full coupling, respectively, such that the tunneling term of lead 1 is $\Gamma c_1^{\dagger}[\gamma_1 - i\beta\gamma_2 + e^{i\delta}(\beta\gamma_3 - i\gamma_4)] + \mathrm{H.c.}$ (and similarly for lead 2 with $\phi$). This is achieved by considering an additional Andreev reflection term of the form $c_1^{\dagger}f_{\rm L(\rm R)}^{\dagger}$. A physical interpretation of $\beta$ is the number of Majoranas residing where each arm of the lead is coupled: one ($\beta=0$) or two ($\beta=1$). The full coupling matrix is therefore
\begin{equation}
  W=\sqrt{\frac{\Gamma}{2\pi}}\left(\begin{array}{cccc}
  1 & i\beta & \beta e^{i\delta} & ie^{i\delta}\\
  1 & -i\beta & \beta e^{-i\delta} & -ie^{-i\delta}\\
  1 & i\beta & \beta e^{i\phi} & ie^{i\phi}\\
  1 & -i\beta & \beta e^{-i\phi} & -ie^{-i\phi}
  \end{array}\right)
\end{equation} in the basis of $\left\{ c_{1}^{\dagger},c_{1},c_{2}^{\dagger},c_{2}\right\} $
and $\left\{ \gamma_{1},\gamma_{2},\gamma_{3},\gamma_{4}\right\}$.
We then calculate the conductance ${\cal G}\left(E\right)$ as a function
of the energy $E$ from the $4\times4$ scattering matrix~\cite{iida_statistical_1990, fisher_relation_1981}
\begin{equation}\label{Weiden}
S\left(E\right)=1-2\pi i W\left(E-H+i\pi W^{\dagger}W\right)^{-1}W^{\dagger},
\end{equation} by using
\begin{equation}\label{eq:G}
{\cal G}\left(E\right)=G_{0}\left[-t_{ee}\left(E\right)+t_{eh}\left(E\right)\right],
\end{equation}
where $t_{ee(eh)}=\left|S_{ee\left(eh\right)}\right|^{2}$ is the
probability for an incoming electron from lead~1 to be transmitted
to lead~2 as an electron (hole), and $G_0=2e^2/h$ is the conductance quantum. We express the conductance as a function of the phase differences $\phi$ and $\delta$. Although it is experimentally challenging to realize two independently tunable phases, $\phi$ and $\delta$, we consider them here as such to fully explore their phase space.

In the topological case, the leads are coupled to a single Majorana at each side, hence $\beta=0$. Considering a
finite temperature $T\gg\Gamma$, we obtain the energy-averaged
conductance approximated up to second order in
$\Gamma/T$ as a function of $\phi$, 
\begin{equation}\label{eq:G_topo}
G\approx -G_{0}\frac{\Gamma}{T}\left(\pi-4\frac{\Gamma}{T}\right)\cos\delta\cos\phi.
\end{equation}
That is, the conductance always peaks at $\phi=\pi$ and $\delta=\pi$, and is zero for $\phi=\pi/2$ and $\delta=\pi/2$.

\begin{figure*}
\centering
\includegraphics[width=\textwidth]{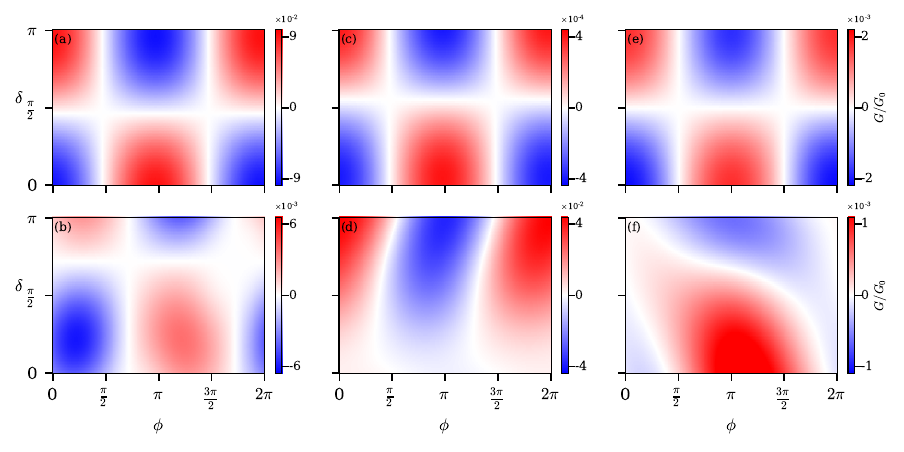}
\caption{Non-local conductance at $E=0$ as a function of the phase differences $\delta$ and $\phi$, calculated for three distinct realizations of TSCs with different physical parameters: (a)~topological and (b)~non-topological (quasi-Majoarana) multi-mode
wires, (c)~topological and (d)~non-topological (Andreev bound state) SNS systems,
and (e)~topological and (f)~non-topological (magnetic defect) three-phase systems. The results demonstrate the robustness of the location of the conductance peak, only in the topological cases, and across all three systems. Notice the different color scales for each panel, due to the different physical parameters of each system.
\label{fig:arms}}
\end{figure*}

On the other hand, in the non-topological case, when all four Majoranas are equally coupled to each lead, forming a full fermion on each side of
the SC ($\beta=1$), $t_{eh}=0$ and the conductance takes the form 
\begin{equation}
G\approx-G_{0}\frac{\Gamma}{T}\left(\frac{\pi}{2}-4\frac{\Gamma}{T}\right)\left[\cos\left(\phi-\delta\right)+2\right].
\end{equation}
This expression depends on $\phi-\delta$ such that it no longer peaks at $\phi=\pi$ and $\delta=\pi$. A smooth transition between the cases is observed as we tune the coupling of the leads to the inner Majorana components by changing $\beta$ (further details on the analytical model are provided in the Supplemental Material).

Therefore, we find a clear difference between the topological and non-topological cases. In the former, when only two Majoranas are present, the interference pattern of the conductance is robust under the change of $\delta$ and $\phi$ and always peaks at the same values. However, in the latter case, the position of the peak depends on $\phi-\delta$.

\section{Numerical model}To corroborate our analytical findings, we numerically test two experimentally realistic schemes with a full tight-binding model~\cite{datta_electronic_1995}. Both share the structure
described above and differ in the specific perturbation used to probe the number
of present Majoranas. The first one follows the underlying scheme
of the analytical model, in which the relative phase
of the incoming electron, $\delta$, is changed. In the second, the low-energy states
are perturbed by a gate potential, affecting the phase acquired by tunneling through them.

In both experimental schemes, we test three different
realizations of TSCs: (i)~a multi-mode superconducting
wire with spin-orbit coupling (SOC) and a Zeeman field~\cite{oreg_helical_2010,lutchyn_majorana_2010}; (ii)~a long phase-biased Josephson junction with a Zeeman field (SNS)~\cite{pientka_topological_2017,hell_two-dimensional_2017,fornieri_evidence_2019,ren_topological_2019,banerjee_local_2023}; (iii)~a double Josephson junction using just phase control (three-phase)~
\cite{lesser_one-dimensional_2022}. For each system, the topological case was compared to cases of non-topological
localized states obtained from magnetic impurities or quasi-Majoranas
 stemming from smooth potentials, similar to Ref.~\cite{vuik_reproducing_2019}.   
In all systems, we denote the long axis as $x$, such that the localized states live around
$x=0$ and $x=L$, and they extend along the width of the system,
the $y$ axis.
As these systems are now spinful, we calculate the conductance as
\begin{equation}
G\left(E\right)=G_{0}\sum_{\sigma\sigma'}\left[-t_{e\sigma,e\sigma'}\left(E\right)+t_{e\sigma,h\sigma'}\left(E\right)\right].
\end{equation}
Here $t_{e\sigma,e\sigma'}$ represents the probability for an incoming electron
with spin $\sigma$ from lead~1 to be transmitted to lead~2 as an
electron with spin $\sigma'$, and similarly with $t_{e\sigma,h\sigma'}$. Summing over both spins species is crucial for obtaining the robust interference pattern shown below; the spin-resolved conductance generally does not exhibit this behavior.

In all systems, we assumed normally distributed, uncorrelated disorder in the chemical potential. Unlike the MZMs, the non-topological localized states are not locked at strictly zero energy, though they are well within the gap. The disorder and perturbation of the chemical potential may alter their energies. Therefore, and following our results above, we assume a finite temperature $T$ such that $T$ is sufficiently larger than the state's energy and yet much smaller than the higher-energy states.

\section{Phase difference in input electron}A relative phase difference can be achieved through a magnetic flux between the input arms or by adjusting the effective arm lengths using, for instance, electrostatic gates. It is worth noting that the former breaks time-reversal symmetry, while the latter does not. However, the distinction between the two becomes evident only in the presence of significant reflections at the electron-splitting junctions. Here and in
the next scheme, we assumed that the reflections are low ($1\%$) to simplify the discussion and highlight the pertinent features. Numerical results for the case of significant reflections are provided in the Supplemental Material. We choose to implement $\delta$ as an arms-length difference and $\phi$ with a magnetic flux.

Figure~\ref{fig:arms} shows the numerical simulation results, exhibiting features similar to our analytical model above. In all the topological cases
(upper panels), the conductance shows a strict cosine dependence
on $\phi$ and $\delta$, consistent with
$G\propto\cos\delta\cos\phi$ in Eq.~\eqref{eq:G_topo}. It always peaks at $\phi=\pi$, symmetrically oscillating around zero. On the other hand, in the non-topological cases
(bottom panels), the peak's position significantly moves around as $\delta$ changes, and the conductance is not centered around zero. 

\section{Gate perturbation}We now present another scheme employing an electrostatic gate as a perturbation, which is a more accessible knob for manipulating the system. The motivation behind this approach lies in the following: The energy of a single isolated Majorana cannot be modified. Thus, a local perturbation applied to an area containing only a single isolated Majorana should have no effect on the interference pattern.

Accordingly, in this scheme, we locally change the chemical potential on one side
of the system with an electrostatic gate. We scan through $\phi$ while maintaining the relative input phase $\delta$ fixed. Consider the general, non fine-tuned
case in which the wavefunctions are complex. When traveling in the
$y$ direction, the acquired phase is roughly related to $e^{ik_{\rm F}L_y}$,
where $k_{\rm F}$ is the Fermi momentum and $L_y$ the width of the system.
Changing the chemical potential leads to a modification in $k_{\rm F}$, denoted as $k_{\rm F}+\Delta k$, affecting the acquired phase. Hence, a substantial perturbation in phase requires $\Delta kL_y\sim1$.
This sets the energy scale for the perturbation, 
\begin{equation}
\Delta\mu=v_{\rm F}\Delta k\sim\frac{v_{\rm F}}{L_y}=E_{\rm T}
\end{equation}
where $v_{\rm F}$ is the Fermi velocity and $E_{\rm T}$ the Thouless energy~\cite{edwards_numerical_1972}.
In this, we also assumed that $\Delta k$ is small compared to $k_{\rm F}$,
which adds a constraint on the system, such that $k_{\rm F}L_y$
is large enough. Otherwise, no phase will be acquired in any case.

Notably, in this scheme, the isolation of an MZM is probed differently than in the previous scheme. In the first scheme, the key question was how many Majoranas are coupled to the leads. However, in the gate perturbation scheme, the critical question is how many Majoranas are affected by the local electrostatic potential change induced by the gate.

\begin{figure}
\centering
\includegraphics[width=\linewidth]{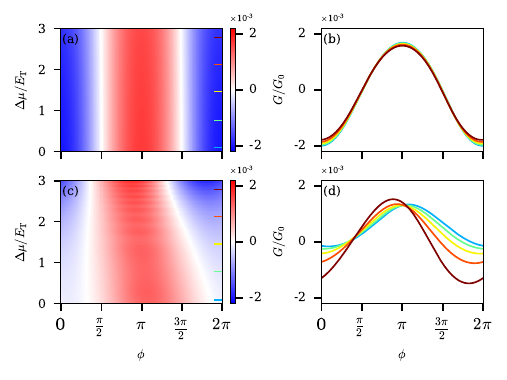}
\caption{Non-local conductance $G$ at $E=0$ as a function of the phase difference $\phi$ and the 
local chemical potential perturbation $\Delta\mu$ in units of $E_{\rm T}$, for (a)~topological and (c)~non-topological multi-mode three-phase systems. (b)~and~(d) show $G$ as a function of $\phi$ for several values of $\Delta\mu/E_{\rm T}$. We set $\delta=0.1\pi$ to account for some constant difference in the length of the input arms.\label{fig:gate}
}

\end{figure}

Figure~\ref{fig:gate} shows the different behavior of topological and non-topological systems under this perturbation. In the topological case, the conductance peak strictly remains at $\phi=\pi$. However, we do observe some slight smooth changes in the peak's amplitude, establishing the fact that the perturbation does act on the system. In contrast, in the non-topological case, neither the peak's position nor its amplitude are maintained. 

\section{Discussion}Our results demonstrate a universal behavior across various topological systems, irrespective of their specific realization. Especially, we observe a consistent peak in conductance at $\phi=\pi$, aligning with the basic interference phenomenon of two coherent signals with zero relative phase. This universality emphasizes the robustness of our proposed experimental scheme in detecting Majorana zero modes.
Notably, this behavior does not manifest when measuring spin-resolved conductance. This observation stems from the encoding of Majoranas across both spins, making their properties inaccessible through measurement of a single spin component. Correspondingly, in systems like the multi-mode wire, where only one spin species is relevant at low energies, the behavior becomes discernible even when considering only the active spin component.

One may intuit that the non-local features demonstrated here arise due to the two Majoranas residing on the edges of the same TSC bulk. Rather, as our analytical model indicates (and corroborated by numerical simulations), the fundamental origin lies in the number of low-energy degrees of freedom being probed. Remarkably, this phenomenon would persist even for Majorana pairs $\gamma_1,\gamma_2$ and $\gamma_3,\gamma_4$ (see Fig.~\ref{fig:system}) localized on the edges of two disconnected TSCs. This also hints that the leads play a dual role, not only probing but also enabling information transfer between spatially separated Majoranas.

A significant advantage of this method is in its ability to leverage both disorder and temperature as advantageous factors. In clean systems, spatial symmetries can mimic the topological protection, obscuring the distinction between non-topological and topological phases. Introducing moderate disorder breaks these spatial symmetries, thus accentuating the differences between the two phases. Additionally, measuring at finite temperature enables avoiding the singularities that emerge at zero energy. 

Unlike inter-Majorana interferometry (in grounded SCs)~\cite{sugeta_enhanced_2023}, our approach does not depend on the exponentially suppressed coupling of the MZMs. The anticipated conductance we predict typically falls between $10^{-3}$ and $10^{-1}$ $G_0$, regardless of the system's length, making it well within experimental reach. This paves the way for the conclusive demonstration of the non-local nature of the Majorana zero modes.

\section{Acknowledgments}We are grateful to Samuel D. Escribano for fruitful discussions. This work was supported by the European Union's Horizon 2020 research and innovation programme (Grant Agreement LEGOTOP No. 788715) and the DFG (CRC/Transregio 183, EI 519/7-1). 

\bibliography{library}

\clearpage
\setcounter{secnumdepth}{2}
\onecolumngrid
\let\section\oldsection

\begin{center}
\Large{\textbf{Supplemental Material}}
\end{center}

\setcounter{equation}{0}
\renewcommand{\theequation}{S\arabic{equation}}
\setcounter{figure}{0}
\renewcommand{\thefigure}{S\arabic{figure}}
\setcounter{section}{0}
\renewcommand{\thesection}{S\Roman{section}}

\section{Analytical Model}
We consider the zero Hamiltonian and the coupling matrix 
\begin{equation}
  W=\sqrt{\frac{\Gamma}{2\pi}}\left(\begin{array}{cccc}
  1 & i\beta & \beta e^{i\delta} & ie^{i\delta}\\
  1 & -i\beta & \beta e^{-i\delta} & -ie^{-i\delta}\\
  1 & i\beta & \beta e^{i\phi} & ie^{i\phi}\\
  1 & -i\beta & \beta e^{-i\phi} & -ie^{-i\phi}
  \end{array}\right),
\end{equation}
as described in the main text. Having obtained the scattering matrix $S(E)$ as in~\eqref{Weiden}, we extract $t_{ee}(E)$ and $t_{eh}(E)$ to calculate the conductance G via \eqref{eq:G} for the case
$\beta=0$:

\begin{align}
t_{ee} & =\left|S_{ee}\right|^{2}=\left|\frac{4\Gamma\cos\left(\frac{\delta-\phi}{2}\right)\left(\Gamma\left(\cos\left(\phi+\delta\right)+1\right)-iE\right)}{8iE\Gamma-\Gamma^{2}\left(\cos2\phi-4\sin\phi\sin\delta+\cos2\delta+6\right)+2E^{2}}\right|^{2}\\
 & =\frac{16\cos^{2}\left(\frac{\delta-\phi}{2}\right)\left(\Gamma{}^{4}\left(\cos\left(\phi+\delta\right)+1\right)^{2}+\Gamma^{2}E^{2}\right)}{\left(\Gamma^{2}\left(\cos2\phi-4\sin\phi\sin\delta+\cos2\delta+6\right)-2E^{2}\right)^{2}+64\Gamma^{2}E^{2}}
\end{align}

and 
\begin{align}
t_{eh} & =\left|S_{eh}\right|^{2}=\left|\frac{4i\Gamma\sin\left(\frac{\phi+\delta}{2}\right)\left(\Gamma\left(\cos\left(\phi-\delta\right)-1\right)+iE\right)}{8iE\Gamma-\Gamma^{2}\left(\cos2\phi-4\sin\phi\sin\delta+\cos2\delta+6\right)+2E^{2}}\right|^{2}\\
 & =\frac{16\sin^{2}\left(\frac{\phi+\delta}{2}\right)\left(\Gamma^{4}\left(\cos\left(\phi-\delta\right)-1\right)+\Gamma^{2}E^{2}\right)}{\left(\Gamma^{2}\left(\cos2\phi-4\sin\phi\sin\delta+\cos2\delta+6\right)-2E^{2}\right)^{2}+64\Gamma^{2}E^{2}},
\end{align}
so we get 
\[
-t_{ee}+t_{eh}=-\frac{8\Gamma^{2}\left[\cos\phi\cos\delta\left(2\left(3\Gamma^{2}+E^{2}\right)+\Gamma^{2}\left(\cos2\phi-4\sin\phi\sin\delta+\cos2\delta\right)\right)\right]}{\left(\Gamma^{2}\left(\cos2\phi-4\sin\phi\sin\delta+\cos2\delta\right)+6\Gamma^{2}-2E^{2}\right)^{2}+64\Gamma^{2}E^{2}}.
\]
Considering a finite temperature $T\gg\Gamma$, integration of the energy-dependent conductance yields
\begin{align}\label{eq:integral}
G & =G_{0}\frac{1}{2T}\intop_{-T}^{T}\left[t_{eh}\left(E\right)-t_{ee}\left(E\right)\right]dE\\
 & =\ensuremath{-\frac{\Gamma}{T}G_{0}\left[\pi+\tan^{-1}\left(\frac{\Gamma}{T}\left(\sin\phi+\sin\delta-2\right)\right)-\tan^{-1}\left(\frac{\Gamma}{T}\left(\sin\phi+\sin\delta+2\right)\right)\right]\cos\delta\cos\phi}
\end{align}
In the limit $\Gamma/T\ll1$, this expression approximates to
\begin{align}
G & \approx-\frac{\Gamma}{T}G_{0}\left(\pi+\frac{\Gamma}{T}\left(\sin\phi+\sin\delta-2\right)-\frac{\Gamma}{T}\left(\sin\phi+\sin\delta+2\right)\right)\cos\phi\cos\delta\\
 & =\ensuremath{-\frac{\Gamma}{T}G_{0}\left(\pi-4\frac{\Gamma}{T}\right)\cos\phi\cos\delta}.
\end{align}
up to 2$^{nd}$ order in $\Gamma/T$.

For the case of $\beta=1$, as mentioned in the main text, we get $t_{eh}=0$ and 
\begin{align}
t_{ee} = \frac{4\Gamma^{2}E^{2}\left(\cos\left(\phi-\delta\right)+2\right)}{\left(2\Gamma^{2}\left(\cos\left(\phi-\delta\right)-1\right)+E^{2}\right)^{2}+16\Gamma^{2}E^{2}}
\end{align}

By integrating around $E=0$ as in \eqref{eq:integral} we get 
\begin{align}
G &=-G_{0}\ensuremath{\frac{\Gamma}{T}\left(\cos\left(\phi-\delta\right)+2\right)} \times \nonumber \\
&\left[\left(\sec\left(\frac{\phi-\delta}{2}\right)+1\right)\tan^{-1}\left(\frac{T}{4\Gamma}\sec^{2}\left(\frac{\phi-\delta}{4}\right)\right)- \right. \nonumber \\
&\left. \left(\sec\left(\frac{\phi-\delta}{2}\right)-1\right)\tan^{-1}\left(\frac{T}{4\Gamma}\csc^{2}\left(\frac{\phi-\delta}{4}\right)\right)\right]
\end{align}

Assuming again $T\gg\Gamma$ this approximates to
\begin{align}
G\approx-G_{0}\frac{\Gamma}{T}\left(\frac{\pi}{2}-4\frac{\Gamma}{T}\right)\left[\cos\left(\phi-\delta\right)+2\right].
\end{align}

up to 2$^{nd}$ order in $\Gamma/T$.

\begin{figure}[h!]
\centering
\includegraphics[scale=0.7]{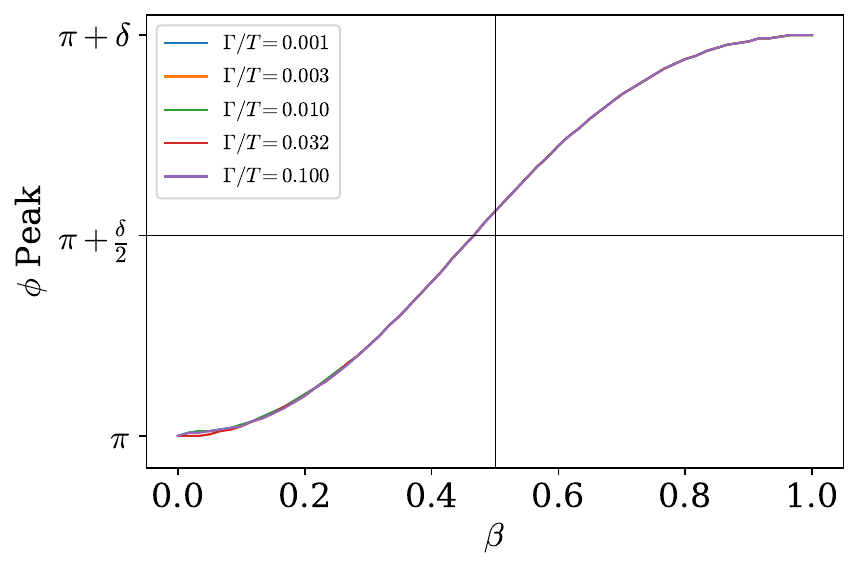}
\caption{The peak of the conductance $G(\phi)$ as a function of $\beta$ for different values of $\Gamma$, the contact energy scale, and $\delta=\pi/4$. In the case of $\beta=0$, where only two Majoranas are present, and the conductance peaks at $\phi=\pi$ and for $\beta=1$ at $\phi=\delta$\label{fig:peak}
}
\end{figure}

\begin{figure}[h!]
\centering
\includegraphics[scale=0.7]{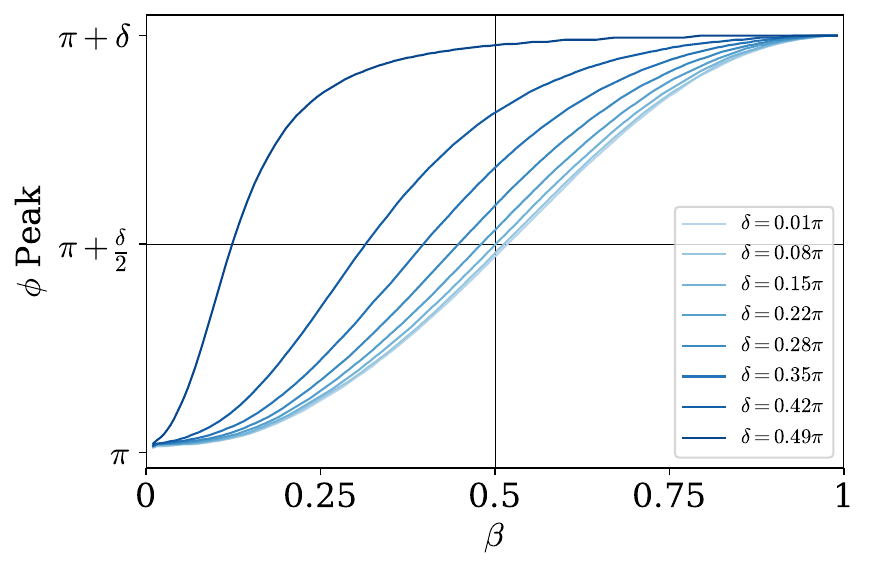}
\caption{The peak of the conductance $G(\phi)$ as a function of $\beta$ for different values of $\delta$. \label{fig:peak_delta}
}
\end{figure}

For the general $\beta$ case we can solve this numerically and plot
the peak of $G\left(\phi\right)$ as a function of $\beta$, as shown in Fig.~\ref{fig:peak}. The resulting functionality is insensitive to the specific value of $\Gamma$ as long as $\Gamma/T\ll1$, but depends on the value of $\delta$ as seen in Fig.~\ref{fig:peak_delta}.
\pagebreak
\section{Numerical Results with Reflection}

\begin{figure}[h!]
\centering
\includegraphics[scale=0.6]{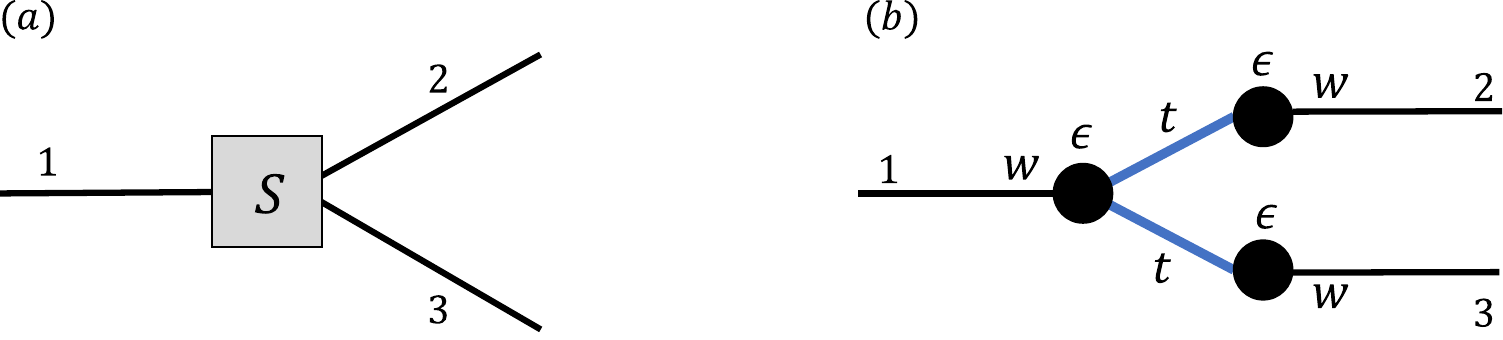}
\caption{Illustration of (a) The scattering matrix S and (b) the tight-binding Hamiltonian describing the splitting of the leads.
\label{fig:splitter}
}
\end{figure}

\begin{figure}[h!]
\centering
\includegraphics[scale=1.2]{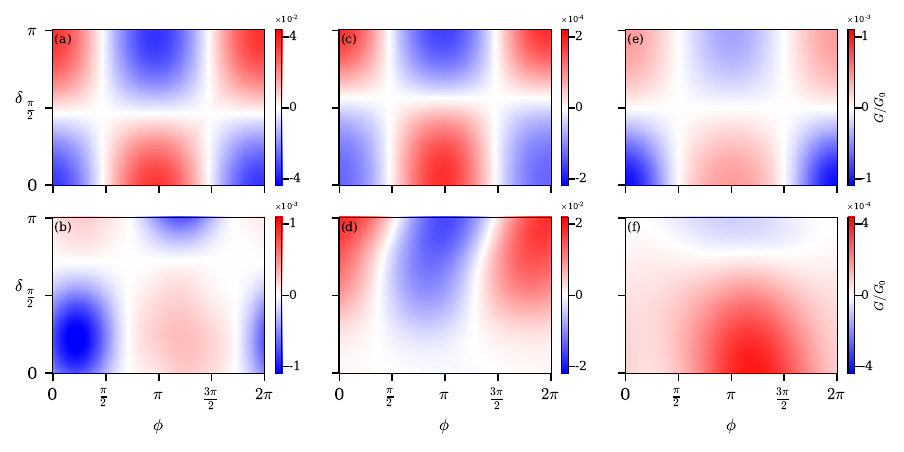}
\caption{Same as Fig.~\ref{fig:arms} from the main text, for the case in which the reflection in the lead splitting is $R=32.5\%$.
\label{fig:arms_ref}
}
\end{figure}

\begin{figure}[h!]
\centering
\includegraphics[scale=1.2]{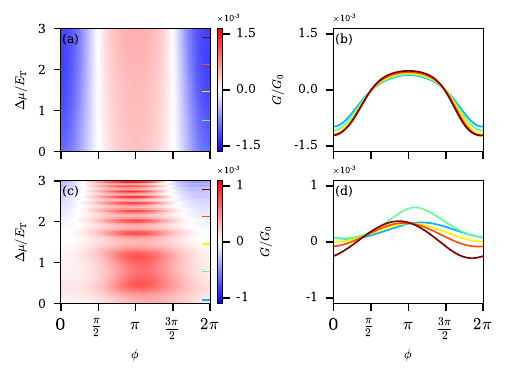}
\caption{Same as Fig.~\ref{fig:gate} from the main text, for the case in which the reflection in the lead splitting is $R=32.5\%$.
\label{fig:gate_ref}
}
\end{figure}

Here, we show the results of our numerical simulation in the presence of higher reflections at the junctions where the normal leads split into two arms.
We model the splitting of the lead with a three-site tight-binding model, with two parameters - $t$ and $\epsilon$ as illustrated in Fig.~\ref{fig:splitter}. Using Eq.~\eqref{Weiden} from the main text, with the coupling strength $w$ the coupling strength we obtain the $3\times3$ scattering matrix $S$. The reflection $R$ is defined as $R=\left|S_{11}\right|^{2}$ which can be controlled by tuning the coupling strength $w$. In the following case, we choose $\epsilon=-1.05t$ and $w=0.35t$ to obtain $R=32.5\%$. As seen in Figs.~\ref{fig:arms_ref},~\ref{fig:gate_ref}, the fundamental robustness of the peak's position is maintained even for significant reflection values.

\pagebreak
\section{Technical Simulation Details}

All energy scales are in units of the hopping element $t$ unless mentioned otherwise,
and lengths in units of number of sites. In all systems $\Delta=0.1$.
$\sigma_{\mu}$ is the standard deviation of the chemical potential
uncorrelated Gaussian disorder. $E_z$ is the Zeeman energy in the $z$ direction. $L$ is the length in $x$ direction, and $W$ width in $y$.
\begin{enumerate}
\item Fig.~\ref{fig:arms}:
\begin{enumerate}
\item Multi-mode wires~\cite{oreg_helical_2010,lutchyn_majorana_2010}:
\begin{enumerate}
\item Topological: $\alpha=0.6$, $L=500$, $W=30$, $\mu=0.5$, $E_{z}=0.63$,
$\sigma_{\mu}=0.1\Delta$, $T=0.03\Delta$. 
\item Non-Topological: $\alpha=0.25$, $L=500$, $W=10$, $\mu=0.5$, $E_{z}=0.3$,
$\sigma_{\mu}=0.1\Delta$, $T=0.03\Delta$. The localized states are
formed with a Gaussian-shaped potential with height $V_{0}=0.5$ and width
of $\sigma=10$ sites, on both edges of the wire.
\end{enumerate}
\item SNS~\cite{pientka_topological_2017,hell_two-dimensional_2017}:
\begin{enumerate}
\item Topological: $\alpha=0.6$, $L=700$, $W_{\rm SC}=12$, $W_{\rm N}=4$, the
phase difference between the SCs is $\Delta\phi=0.8\pi$, $\mu=0.5$,
$E_{z}=0.15$, $\sigma_{\mu}=0.05\Delta$, $T=0.005\Delta$. 
\item Non-Topological: $\alpha=2$, $L=700$, $W_{\rm SC1}=12$, $W_{\rm SC1}=10$,
$W_{\rm N}=2$, the phase difference between the SCs is $\Delta\phi=\pi$,
$\mu=0.5$, $E_{z}=0$, $\sigma_{\mu}=0.05\Delta$, $T=0.007\Delta$. 
\end{enumerate}
\item Three-phase~\cite{lesser_one-dimensional_2022}:
\begin{enumerate}
\item Topological: The n.n.n hopping element is $t'=0.94$, $\alpha=1.85$,
$L=1000$, for the outer SC $W_{\text{SC-O}}=12$, and for the middle one
$W_{\text{SC-M}}=10$, $W_{\rm N}=3$, $\phi_{1}=\pi,\phi_{2}=0,\theta=0.4\pi$,
$\mu=0.6$, $\sigma_{\mu}=0.1\Delta$, $T=0.03\Delta$. 
\item Non-Topological: Same as the topological case but $\phi_{1}=\pi,\phi_{2}=0,\theta=0$, $\mu=0.6$, $T=0.025\Delta$. The magnetic impurities
are placed across $y$, and extend only to a single site in $x$ on both edges of the
system, with strength $J=3$.
\end{enumerate}
\end{enumerate}
\item Fig.~\ref{fig:gate}: Same as mentioned above for Fig.~\ref{fig:arms} in the three-phase
case. The electrical gate $\Delta\mu$ acts on a $30\times W$ area
at the left edge of the system.
\end{enumerate}

\end{document}